\documentstyle[12pt]{article}

\def\1ad{\mbox{\normalsize $^1$}}
\def\2ad{\mbox{\normalsize $^2$}}
\def\3ad{\mbox{\normalsize $^3$}}
\def\4ad{\mbox{\normalsize $^4$}}
\def\5ad{\mbox{\normalsize $^5$}}
\def\6ad{\mbox{\normalsize $^6$}}
\def\7ad{\mbox{\normalsize $^7$}}
\def\8ad{\mbox{\normalsize $^8$}}
\def\makefront{\vspace*{1cm}\begin{center}
\def\newtitleline{\\ \vskip 5pt}
{\Large\bf\titleline}\\
\vskip 1truecm
{\large\bf\authors}\\
\vskip 5truemm
\addresses
\end{center}
\vskip 1truecm
{\bf Abstract:}
\abstracttext
\vskip 1truecm}
\newcommand{\eqn}[1]{(\ref{#1})}
\newcommand{\ft}[2]{{\textstyle\frac{#1}{#2}}}
\newsavebox{\uuunit}
\sbox{\uuunit}
    {\setlength{\unitlength}{0.825em}
     \begin{picture}(0.6,0.7)
        \thinlines
        \put(0,0){\line(1,0){0.5}}
        \put(0.15,0){\line(0,1){0.7}}
        \put(0.35,0){\line(0,1){0.8}}
       \multiput(0.3,0.8)(-0.04,-0.02){12}{\rule{0.5pt}{0.5pt}}
     \end {picture}}
\newcommand {\Cbar}
    {\mathord{\setlength{\unitlength}{1em}
     \begin{picture}(0.6,0.7)(-0.1,0)
        \put(-0.1,0){\rm C}
        \thicklines
        \put(0.2,0.05){\line(0,1){0.55}}
     \end {picture}}}

\newcommand{\Ka}{K\"ahler}

\def\IP{\relax{\rm I\kern-.18em P}}
\def\IH{\relax{\rm I\kern-.18em H}}
\def\IC{\Cbar}

\font\cmss=cmss10 \font\cmsss=cmss10 at 7pt
\def\ZZ{\relax\ifmmode\mathchoice
{\hbox{\cmss Z\kern-.4em Z}}{\hbox{\cmss Z\kern-.4em Z}}
{\lower.9pt\hbox{\cmsss Z\kern-.4em Z}}
{\lower1.2pt\hbox{\cmsss Z\kern-.4em Z}}\else{\cmss Z\kern-.4em
Z}\fi}

\csname @addtoreset\endcsname{equation}{section}

\font\cmss=cmss10 \font\cmsss=cmss10 at 7pt

\def\inbar{\vrule height1.5ex width.4pt depth0pt}
\def\IC{\relax\,\hbox{$\inbar\kern-.3em{\rm C}$}}
\def\IG{\relax\,\hbox{$\inbar\kern-.3em{\rm G}$}}
\def\IB{\relax{\rm I\kern-.18em B}}
\def\ID{\relax{\rm I\kern-.18em D}}
\def\IL{\relax{\rm I\kern-.18em L}}
\def\IF{\relax{\rm I\kern-.18em F}}
\def\IH{\relax{\rm I\kern-.18em H}}
\def\II{\relax{\rm I\kern-.17em I}}
\def\IN{\relax{\rm I\kern-.18em N}}
\def\IP{\relax{\rm I\kern-.18em P}}
\def\IQ{\relax\,\hbox{$\inbar\kern-.3em{\rm Q}$}}
\def\bfzero{\relax\,\hbox{$\inbar\kern-.3em{\rm 0}$}}
\def\IR{\relax{\rm I\kern-.18em R}}
\def\ZZ{\relax\ifmmode\mathchoice
{\hbox{\cmss Z\kern-.4em Z}}{\hbox{\cmss Z\kern-.4em Z}}
{\lower.9pt\hbox{\cmsss Z\kern-.4em Z}}
{\lower1.2pt\hbox{\cmsss Z\kern-.4em Z}}\else{\cmss Z\kern-.4em
Z}\fi}

\begin {document}

\begin{titlepage}
\begin{flushright} KUL-TF-98/4\\ DFTT-1/98\\
 IFUM/604-FT \\ hep-th/9801112
\end{flushright}
\begin{center}
{\Large\bf A detailed case study of the rigid limit in Special K\"ahler
geometry using $K3$.
}
\vskip 0.3cm
{\large {\sl }}
{\bf   Marco Bill\'o$^1$,  Frederik Denef $^{1,+}$, \\ 
Pietro Fr\`e$^2$,
Igor Pesando$^2$, Walter Troost$^{1,*}$,\\ 
Antoine Van Proeyen $^{1,\dagger}$ and  Daniela Zanon$^3$ } \\
\vfill
{\small
$^1$ Instituut voor theoretische fysica, \\
Katholieke Universiteit Leuven, B-3001 Leuven, Belgium\\
 $^2$  Dipartimento di Fisica Teorica dell' Universit\`a,\\
                 via P. Giuria 1,
                I-10125 Torino, Italy\\
 $^3$ Dipartimento di Fisica dell'Universit\`a di Milano and\\
INFN, Sezione di Milano, via Celoria 16,
I-20133 Milano, Italy. }
\end{center}
\vfill
\begin{center}   {\bf
To be published in the proceedings of\\ the 31$^{st}$ International
Symposium Ahrenshoop on the Theory of Elementary Particles.\\
Buckow, Brandenburg, Germany,
sept.2-6, 1997.}\\ Talk presented by A.V.P.
\end{center}
\vfill
\begin{center}
{\bf ABSTRACT}
\end{center}
\begin{quote}
This is a r\'esum\'e of an extensive investigation of some examples
in which one obtains the rigid limit of $N=2$ supergravity by means
of an expansion around singular points in the moduli space of a
Calabi-Yau 3-fold. We make extensive use of the $K3$ fibration of the
Calabi-Yau manifolds which are considered. At the end the fibration
parameter becomes the coordinate of the Riemann surface whose moduli
space realises rigid $N=2$ supersymmetry.
\vskip 2.mm
 \hrule width 5.cm
\vskip 2.mm
{\small
Work supported by
the European Commission TMR programme ERBFMRX-CT96-0045,
in which D.Z. is associated to U. Torino.\\
\noindent $^+$ Aspirant FWO, Belgium \\
\noindent $^*$ Onderzoeksleider FWO, Belgium \\
\noindent $^\dagger$ Onderzoeksdirecteur FWO, Belgium \\
 }
\end{quote}
\end{titlepage}
\def\titleline{ A detailed case study of the rigid limit in \newtitleline
Special K\"ahler geometry using $K3$. }
\def\authors{  Marco Bill\'o \1ad,
Frederik Denef \1ad,
Pietro Fr\`e \2ad, \\
Igor Pesando \2ad,  Walter Troost \1ad,\\
Antoine Van Proeyen \1ad
and  Daniela Zanon \3ad }
\def\addresses{  \1ad
Instituut voor theoretische fysica, \\
Katholieke Universiteit Leuven, B-3001 Leuven, Belgium\\
\2ad Dipartimento di Fisica Teorica dell' Universit\`a,\\
                 via P. Giuria 1,
                I-10125 Torino, Italy\\
\3ad Dipartimento di Fisica dell'Universit\`a di Milano and\\
INFN, Sezione di Milano, via Celoria 16,
I-20133 Milano, Italy. }
\def\abstracttext{
This is a r\'esum\'e of an extensive investigation of some examples
in which one obtains the rigid limit of $N=2$ supergravity by means
of an expansion around singular points in the moduli space of a
Calabi-Yau 3-fold. We make extensive use of the $K3$ fibration of the
Calabi-Yau manifolds which are considered. At the end the fibration
parameter becomes the coordinate of the Riemann surface whose moduli
space realises rigid $N=2$ supersymmetry.
}
\makefront
The vector multiplet of $N=2$, $d=4$ supersymmetry often occurs in
the study of string dualities. This is related to the fact that $N=2$
is the minimal supersymmetry to connect the scalars to the vectors,
which in four dimensions have duality transformations between their
electric and magnetic field strengths. These transformations are in a
symplectic group, and therefore the structure of the manifold of the
scalars also inherits this symplectic structure. The resulting
geometry of these complex scalars is denoted as `special \Ka\
geometry', and exists as well for rigid as for local supersymmetry.
In both cases this geometry can be realised on complex structure moduli spaces: for
rigid geometry on moduli spaces of a class of Riemann
surfaces (RS),  for local supersymmetry (supergravity) on
 moduli spaces of Calabi-Yau 3-folds (CY).
The relevant objects which build the supersymmetric actions
are `periods',  of 1-forms
over 1-cycles in the case of RS, of 3-forms
over 3-cycles in the case of CY. These forms depend on the moduli which are identified
with the scalar fields of the supersymmetric theory.

Many relevant CY manifolds in string theory are $K3$-fibrations, and
we will restrict to these. That means that the manifold can be
described as a (complex 2-dimensional) $K3$ surface for which the
moduli depend on the moduli of the CY but also on a complex variable,
denoted as $\zeta$, which can be viewed as the third complex
dimension of the CY. Thus for any fixed value of $\zeta$, the CY is
a $K3$-manifold. As such e.g. the unique CY (3,0)-form will be
represented as $d\zeta\wedge \Omega^{(2,0)}$, where the latter is the
$(2,0)$-form of the $K3$. The 3-cycles of the CY manifolds on the
other hand can be obtained in 2 different ways. One can consider a
path between two (singular) points in the $\zeta$-base space where the same
$K3$ cycle vanishes. Combining the 2-cycle above these points leads
to one type of CY 3-cycles. Another one can be constructed by
considering in the base space a loop around such a singular point
combined again with the $K3$ 2-cycle which vanishes at that point.
We can often calculate the integral over the $K3$ 2-cycles. Then
the CY period reduces to the integral of a 1-form over the 1-cycle
in the $\zeta$ plane. The latter is not yet a Riemann surface
however.

As already mentioned, the CY moduli space has singular points where
cycles degenerate. We consider an expansion around certain singular
points. The expansion is as well an expansion in moduli space as in
the CY coordinates. The CY moduli $z^\alpha$ become in this way a
function of the expansion parameter $\epsilon$ and variables
$u^i$, which will become the moduli of a Riemann surface. In this way
the local geometry is expanded so that a rigid special \Ka\
geometry remains. In this expansion the $K3$ manifold reduces to an
ALE manifold. By performing the 2-dimensional integrals,
the periods of the CY reduce to periods of an element of this class
of Riemann surfaces. We have made an expansion from a supergravity
model to a rigid supersymmetric one.

In supergravity a rigid limit is not defined a priori. In
the present framework this is reflected in that different singular
points may give rise to different rigid limits. In \cite{klemm} a
procedure was set up to reduce the CY to an ALE manifold, leading to
such rigid limits. See \cite{lectures} for further references.
Rather than using this reduction, we computed \cite{cyk3art} all the
periods in the picture of the $K3$-fibration. This shows explicitly
how the full supergravity model approaches its rigid limit. Some
cycles which do not occur in the ALE manifolds lead to periods whose
contribution give in the limit $\epsilon\rightarrow 0$ (infinite
Planck mass) a diverging renormalisation of the rigid \Ka\ potential.
Thus these renormalisation effects are included in our computation.
 For full references see \cite{cyk3art}.
\section{Special \Ka\ geometry and CY moduli spaces}
First we summarise the relevant geometric concepts, both for
the rigid and for the local case. We consider
symplectic vectors $V(u)$ (rigid), resp. $v(z)$ (local)
which are holomorphic functions of $r$, (resp. $n$)
complex scalars $\{u^i\}$ (resp. $\{z^\alpha\}$).
These are
$2r$-vectors for the rigid case (in correspondence with the electric
and magnetic field strengths), and $2(n+1)$-vectors in the local
case (because in that case there is also the graviphoton). A
symplectic inner product is defined as
\begin{equation}
\left< V,W \right>=V^T\, Q^{-1}\, W\ ;\qquad
\left< v,w \right>=v^T\, q^{-1}\, w\ ,     \label{symplinner}
\end{equation}
where $Q$ (and $q$) is a real, invertible, antisymmetric matrix (we wrote $Q^{-1}$
in \eqn{symplinner} in view of the meaning which $Q$ will get in the
moduli space realisations).

The \Ka\ potential is respectively for the rigid and local manifold
\begin{equation}
K(u,{\bar u})=i\left< V(u),\bar V(\bar u)\right>\ ;     \qquad
{\cal K}(z,{\bar z})=-\log(-i\left< v(z),\bar v(\bar z)\right>)\ . \label{Kahlerpotentials}
\end{equation}
In the rigid case there is a rigid invariance $V\rightarrow e^{i\theta}V$,
but in the local case there is even a symmetry with a holomorphic
function: $v(z)\rightarrow e^{f(z)}v(z)$, because this gives a \Ka\
transformation ${\cal K}(z,\bar z)\rightarrow {\cal K}(z,\bar z) - f(z) -\bar f(\bar z)$.
Because of this local symmetry we have to introduce covariant
derivatives $
{\cal D}_{\bar \alpha}v=\partial _{\bar \alpha}v= 0
$ and $
{\cal D}_\alpha v=\partial_\alpha v +\left( \partial_\alpha {\cal K}\right) v
$
(There exists also a more symmetrical formulation).
In any case we
still need one more constraint (leading to the `almost always'
existence of a prepotential), which is for rigid, resp. local
supersymmetry:
\begin{equation}
\left< \partial_i V,\partial_j V\right>=0\ ;  \qquad
\left< {\cal D}_\alpha v,{\cal D}_\beta v\right>=0\ .
\end{equation}
There are further global requirements;
 for an exact formulation we refer to \cite{whatskg}.
\par
Local special \Ka\ geometry is realised in moduli spaces of CY
manifolds. Consider a CY manifold with $h^{21}=n$. It
has $n$ complex structure moduli to be identified with the
complex scalars $z^\alpha$. There are $2(n+1)$ 3-cycles $c_\Lambda$,
whose intersection matrix will be identified with the symplectic metric
$q_{\Lambda\Sigma}=c_\Lambda\cap  c_\Sigma$. One
identifies  $v$ with the `period' vector formed by integration of the $(3,0)$
form over the $2(n+1)$ cycles:
\begin{equation}
v=\int_{c_\Lambda}\Omega^{(3,0)}\ ;\qquad
{\cal D}_\alpha v=\int_{c_\Lambda}\Omega^{(2,1)}_{(\alpha)}\ .
\end{equation}

Rigid special \Ka\ geometry is realised in moduli spaces of RS.
A RS of genus $g$ has $g$ holomorphic $(1,0)$ forms.
Now in general we need a family of Riemann surfaces with $r$ complex
moduli $u^i$, such
that one can isolate $r$ (1,0)-forms which are the derivatives of a
meromorphic 1-form $\lambda$ up to a total derivative:
\begin{equation}
\gamma_i=\partial_i\lambda +d\eta_i\
;\qquad\alpha=1,\ldots ,r\leq g\ .
\end{equation}
Then one should also identify $2r$ 1-cycles $c_A$ forming a
complete basis for the cycles over which the integrals of $\lambda$
are non-zero. We  identify $ V=\int_{c_A}  \lambda  $,
but it should be clear  that all this is
much less straightforward then in the CY moduli space.
\section{Description of a Calabi-Yau moduli space}
We present here the description of one of the examples which we use
in \cite{cyk3art}, i.e. a CY space with $n=h^{(12)}=3$. First one
introduces a complex 4-dimensional {\em weighted projective space}
in which points are equivalence classes
$ (x_1,x_2,x_3,x_4,x_5)\sim(\lambda x_1,\lambda x_2,\lambda^2 x_3,
\lambda^8 x_4,\lambda^{12} x_5)$. 
The CY manifold $X_{24}[1,1,2,8,12]$ is a 3-dimensional
submanifold of this projective space, determined by a {\em polynomial
equation} $W=0$, of degree 24. The manifold which we use, $X^*_{24}[1,1,2,8,12]$,
has {\em global identifications}
\begin{eqnarray}
&& x_j \simeq \exp(n_j \frac{2 \pi i}{24}) \, x_j\nonumber\\
&&(n_1,n_2,n_3,n_4,n_5)= m_1 (1,-1,0,0,0) +m_2(-1,-1,2,0,0) \ ,
\label{identif}\end{eqnarray}
where $m_i\in \ZZ$. The most general polynomial of degree 24 which
is invariant under these identifications depends on 11 parameters,
i.e. moduli of the CY
manifold. However, they are not independent: there are
still {\em compatible redefinitions}  of the $x$
variables, i.e. compatible with \eqn{identif} and the weights.
This leaves at the end in this example 3 independent moduli. We learned
the advantages of not restricting immediately to one gauge
choice in this moduli space.

Now {\em the $K3$ fibration} is exhibited by performing the change of
variables
\begin{equation}
x_0=x_1x_2\ ;\qquad \zeta=(x_1/x_2)^{24}\ .  \label{defx0zeta}
\end{equation}
We take a partial gauge choice with one remaining scale invariance,
corresponding to a rescaling of $x_0$.
Then the polynomial looks like
\begin{equation}
W= \ft{1}{12}B' \, x_0^{12} +  \ft{1}{12} x_3^{12}
+\ft{1}{4}x_4^3 +\ft{1}{2} x_5^2 - \psi_0 \left( x_0 x_1 x_3 x_5
\right)
- \ft{1}{6}  \psi_1\left(x_0 x_3 \right)^6\ , \label{K31146}
\end{equation}
with
\begin{equation}
B'=\ft12 B\left(\zeta+\zeta^{-1}\right)-\psi_s \label{B'3mod}\ .
\end{equation}
We thus have a description as a $K3$ manifold $X^*_{12}[1,1,4,6]$, with
a projective moduli space $\{B',\psi_0,\psi_1\}$. Here $B'$
is a function of moduli $B$ and $\psi_s$ of the CY manifold
and contains
the dependence on the base of the $K3$ fibration $\zeta$.
\par
The manifolds are singular when simultaneously $W=0$ and $dW=0$. For
the $K3$ this happens for
\begin{equation}
a)\ B'=(\psi_1+\psi_0^6)^2\ ;  \qquad
b)\  B'=\psi_1^2  \ ;  \qquad
c)\ B'=0\ .
\end{equation}
The singularities then occur for a specific point on $K3$. The cases
a) and b) are $A_1$-type singularities. They coincide if $\psi_0^6=0$,
in which case the singularity
becomes of type $A_2$, and if $\psi_0^6=-2\psi_1$, in which case the singularity
becomes of type $A_1\times A_1$. We will concentrate here on the
first possibility: the $A_2$ singularity.
\par
For the CY to be singular, we should also have that the derivative of
$W$ with respect to $\zeta$ is zero, which leads to
$\frac{\partial B'}{\partial \zeta}=0$, satisfied for all $\zeta$ if
$B=0$. So we have a full $\IP^1$ of singularities.
In the CY moduli space, an appropriate expansion is obtained by taking
\begin{equation}
B=12\epsilon\ ; \qquad \psi_1^2=-\psi_s +\epsilon u_1\ ;
\qquad (\psi_1+\psi_0^6)^2= -\psi_s +\epsilon
( u_1+2 u_0^6) \ .\label{expmoduli}
\end{equation}
After a corresponding expansion of the variables of the
CY space around its singular point, the polynomial reduces to one
defining an ALE manifold of type $A_2$, with moduli $u_1$ and $u_2$.
\section{Periods, monodromies, intersection matrix}
The main work is  to obtain the periods for the $K3$ fibre, after
which the CY periods remain as integrals over 1-cycles. First we use
the {\em Picard-Fuchs equations}, which are differential equations for
the integrals of the $(2,0)$ form over the 2-cycles. It turns out that working
in the enlarged moduli space (where gauges are not yet fixed) simplifies the
derivation of these equations, using toric geometry in disguise
without introducing all the formalism, and avoiding its
higher order differential equations.
The independent solutions give a basis for the periods. The periods
are functions of the moduli appearing in the
polynomial, which have branch points in singular points,
 and we have to choose the position of the cuts in this moduli space.
We  choose a basis of solutions in one sheet of this
moduli space. Continuing the periods around such singular points, we
cross the cuts, and arrive to the same values of the
moduli. Reexpressing the analytically continued
periods in the previously chosen basis gives rise to the
{\em monodromy matrices}. A generic basis of solutions to the
differential equations does, however, not correspond to integrals
over integer cycles. Therefore we need also supplementary methods.
\par
In some examples we can {\em integrate over one cycle and
analytically continue.} We
start by integrating the $(2,0)$ form over a cycle which is known in the
neighbourhood of the `large complex structure' singular point. Then
this period is analytically continued to other regions. By following
its analytic continuation we also obtain the other periods. Because
we start here from an integral cycle, we obtain the monodromy matrices in an
integral basis.
\par
In the example described above, the strategy
which we plan to use for CY, can already be used for the
$K3$ periods themselves. Indeed in this case the $K3$ itself is a
{\em torus fibration.} The forms and cycles can be decomposed in
forms and cycles on the torus, fibred over a $\IP^1$.
It has the advantage that we start from the torus, where we know
already a basis of cycles and its intersection matrix.

The result is that we find expressions corresponding to 4 $K3$-cycles
of which one vanishes at singularity a), called $v_\alpha$, one at b), called
$v_\beta$, and two vanish at c), which we will call $t_\alpha$ and
$t_\beta$.
The points of singularity of the $K3$ manifold each occur at two points in the
$\zeta$-plane, one inside the circle $|\zeta|=1$, and one outside. We will take
the cuts from the former to $\zeta=0$, and from the latter to $\infty$. We can
then construct 4 CY cycles by taking the paths in $\zeta$ between the two
points with the same vanishing $K3$-cycle and combining these with the
corresponding 2-cycle in $K3$. These are called $V_{v_\alpha}$, $V_{v_\beta}$,
$V_{t_\alpha}$ and $V_{t_\beta}$. On the other hand we can combine the circle
$|\zeta|=1$ with the 4 $K3$ cycles, obtaining the CY-cycles $T_{v_\alpha}$, $T_{v_\beta}$,
$T_{t_\alpha}$ and $T_{t_\beta}$.
\section{The rigid limit}
Considering then again the expansion of the moduli as in
\eqn{expmoduli}, we see that \eqn{B'3mod} implies that the singularities are in
leading order of $\epsilon$ at
\begin{eqnarray}
&&v_\alpha\ : \ \ft12(\zeta+{\zeta}^{-1})= \ft1{12}( u_1+2 u_0^6) \
;\qquad
v_\beta \ : \ \ft12(\zeta+{\zeta}^{-1})=\ft1{12} u_1\nonumber\\
&&t_\alpha ,t_\beta\ : \ \ft12(\zeta+{\zeta}^{-1})=\frac{1}{12\epsilon}\psi_s\ .
\end{eqnarray}
The former thus keep their position, while the latter move to $\zeta=0$ and
$\zeta=\infty$ when $\epsilon\rightarrow  0$. The cycles $V_{t_\alpha}$
and $V_{t_\beta}$ thus become infinitely stretched, and the corresponding
periods will get a $\log \epsilon$ dependence. The dependence on $\epsilon$ is
obtained from studying the $\epsilon$-monodromies.

By a complex basis transformation one can isolate the different types
of small $\epsilon$ behaviour of the periods, and one rewrites the
period vector $v$ as
\begin{equation}
v=v_0(\epsilon)+ \epsilon^{1/3} v_1(u) + v_2(\epsilon)\ .
\end{equation}
The relevant term will be the $v_1$ term, which has only 4 non-zero
components, corresponding to the cycles
$V_{v_\alpha}$, $V_{v_\beta}$, $T_{v_\alpha}$ and $T_{v_\beta}$, i.e. related
to the singularities which remain at finite $\zeta$ in the limit.
$v_0$ is independent of the moduli. It contains 2 non-zero
components, one of which starts with a constant, and the other one
has a logarithmic dependence alluded to above. Finally
$v_2(\epsilon)$ has as lowest order terms with $\epsilon^{2/3}$ and
$\epsilon^{2/3}  \log \epsilon$. These appear in the two remaining components of $v$.
The intersection matrix  is in this basis (complex antihermitian, not an integral basis)
block diagonal in the mentioned $4+2+2$ components. For
the \Ka\ potential this leads to
\begin{eqnarray}
{\cal K} &=& -\log \left(  -i < v,\bar  v >\right) \nonumber\\
&=&  -\log \left(  -i < v_0(\epsilon),\bar  v_0(\epsilon) >-i |\epsilon|^{2/3}
<v_1(u),\bar v_1(u)>+ R(\epsilon,u,\bar u)\right) \\
&\approx & -\log \left(  -i < v_0(\epsilon),\bar  v_0(\epsilon) >\right) -
\frac{|\epsilon|^{2/3}}{< v_0(\epsilon),\bar  v_0(\epsilon) >}<v_1(u),\bar v_1(u)>
+ \ldots \ ,  \nonumber
\end{eqnarray}
where $R(\epsilon,u,\bar u)$ are higher order terms. The first term
in the final expression is irrelevant, not depending on the
moduli. The second one is with a diverging renormalisation (produced by the
cycles which disappear in the ALE limit) of the
form which it should have for a rigid special \Ka\ manifold: the first of
\eqn{Kahlerpotentials}. Moreover it coincides with the one the $SU(3)$
Seiberg--Witten Riemann surface.

\paragraph{Acknowledgments.}\par
\noindent
Work supported by
the European Commission TMR programme ERBFMRX-CT96-0045,
in which D.Z. is associated to U. Torino. F.D., W.T. and A.V.P. thank
their employer, the FWO Belgium, for financial support.

\end{document}